\documentclass[preprint,12pt]{elsarticle}

\usepackage{graphicx}
\usepackage{amssymb}

\usepackage{lineno}

\usepackage{notoccite}
\begin{document}



\title{Computational modeling of perfusion in dense solid tumors}



\author{ Mohammad Mehedi Hasan Akash$^{1}$,  Nilotpal Chakraborty$^{2}$, and Saikat Basu$^{1\,*}$}
\vspace{0.5cm}

\address{$^1$~Department of Mechanical Engineering, South Dakota State University, Brookings, SD 57007, United States\\
        $^2$~Department of Mechanical Engineering, National Institute of Technology, Tiruchirappalli, Tamil Nadu 620015, India
	     }
	     
\ead{Saikat.Basu@sdstate.edu}

\begin{abstract}
Tracking and characterizing the blood uptake process within solid pancreatic tumors and the subsequent spatio-temporal distribution of red blood cells  are critical to the clinical diagnosis of the cancer. This systematic computational study of physical factors, affecting the percolation and penetration of blood into a solid tumor, can assist in the development of a new objective clinical diagnosis approach and a framework for personalized targeted drugs.
\end{abstract}


\maketitle

\section{Introduction}
Pancreatic cancer is one of the most lethal forms of human cancer and caused over 430,000 deaths globally
in 2018 alone\cite{rawla2019wjo}, and it typically results in more than 30,000 deaths yearly in the United States\cite{li2004lancet}. Crucially, dense solid tumors are associated with over 85 percent of cancers\cite{shemi2015ot}, including pancreatic cancer. Such tumors demonstrate high resistance to diffusive transport\cite{basu2018ijnmbe}, and the ambient scarcity of blood and lymphatic flows hinders convection.These flow physics nuances present a challenge for blood to perfuse into the tumor volume\cite{chim2018cobe}; this despite the enhanced permeability and retention effect often seen during nanoparticle delivery to tumor tissues\cite{maeda2012jcr}.Accurate quantification of the spatio-temporal distribution of particulates in the blood channel and tumor region, driven by blood flow is critical to clinical diagnosis of the cancer\cite{desposito2018nature}. However, the pathophysiology of diseased tissues has a chaotic interaction with the blood uptake process\cite{soltani2011plos}, and on a macro-scale, can change across different subjects and for variabilities in tumor topography \cite{zhang2013bba}.

Note that the computational tools used in this study have been used by our team before while tracking biomedical transport in the respiratory physiology \cite{basu2020scirep, basu2018ijnmbe, perkins2018ohns, tracy2019ifar, farzal2019ifar}.

\subsection{Complexity in modeling particulate transport in blood}
A critical review of the biomedical literature shows that single-phase CFD models, which treat blood as a homogeneous fluid\cite{tracy2019ifar}, cannot provide the information on hemodynamic factors needed to quantify the interactive effects and the spatial distribution of particulate matter suspended in blood. In an actual scenario, due to high cell number and particulate nature of RBCs, blood exhibits non-Newtonian rheology.\cite{basu2020scirep} The spatial variation of blood constituents such as plasma, RBCs and leukocytes may vary with disturbed flows inside the tumor region for concentrated suspensions.\cite{kimbell2019lsm} In such flows, RBCs collide with one another as a result of relative motion between them and aggregate to form larger RBCs. This phenomenon, knows as RBC aggregation is just one of the inter particle phenomena of blood flow, which determine the rheological behavior of blood and hence affect particulate distribution.\cite{perkins2018ohns}Hence deploying a robust CFD model is  essential to capture these kind of phenomena. In this paper, we start with a multiphase non-Newtonian CFD model for dilute suspension blood flow with constant flow rate and validate against experimental results available in literature. Later, the same model is extended to include time varying flow rate representing actual systole and diastole phases of a heart cycle for quantifying blood particulate concentrations in an artery with tumor vasculature.

\section{Part 1:~Validation of multiphase CFD model with experimental results}
\subsection{Methods}
\subsubsection{Geometry and mesh}
Karino and GoldSmith\cite{karino1977flow} have reported disturbed blood flow in the vortex region formed at the sudden concentric expansion of a 151 um into a 504 um diameter glass tube. Experimental results are available for dilute suspension of hardened human RBCs in water as well as concentrated suspension of RBCs in plasma. For our numerical validation of the multiphase model, we consider the dilute suspension flow in water. To study the concentrated suspension flow of RBCs in plasma for an artery with tumor, we simulated pulsatile flow. The geometry and mesh for the validation part is shown in Fig.1. Table 1 shows the quality of the mesh in terms of standard mesh parameters.
\begin{figure}
 \includegraphics[scale=1.2]{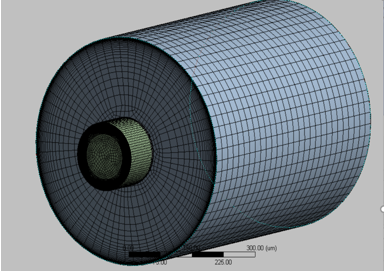}
 \includegraphics[scale=0.5]{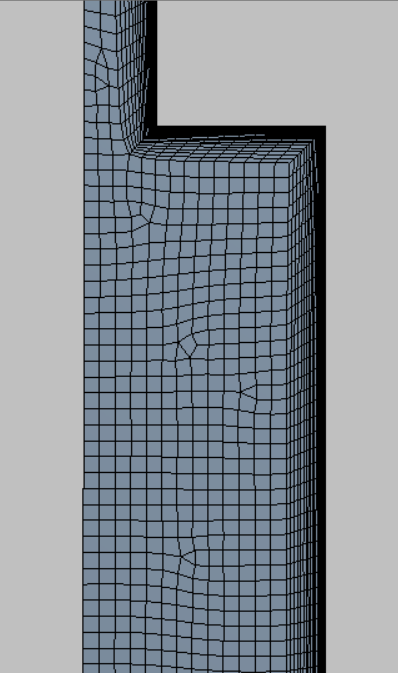} 
 \caption{(a) Geometry used in Karino's experiment (b) Corresponding mesh }
\end{figure}
The above geometry is adjusted for two different Reynolds numbers, Re = 12.2 and Re = 37.8 to replicate Karino and GoldSmith's experiments. For Re = 12.2, the smaller tube is 120 um long and the bigger tube 500 um. For Re = 37.8, the smaller tube is 350 um long and the bigger tube 750 um. Note that in both cases, an extra length l = 0.06*Re*(diameter) is added to the smaller tube to ensure that the flow is fully developed before entering the bigger tube\cite{jung2006hemodynamic}. Fig 2 shows the above concept.
\begin{figure}
\includegraphics[scale=0.5]{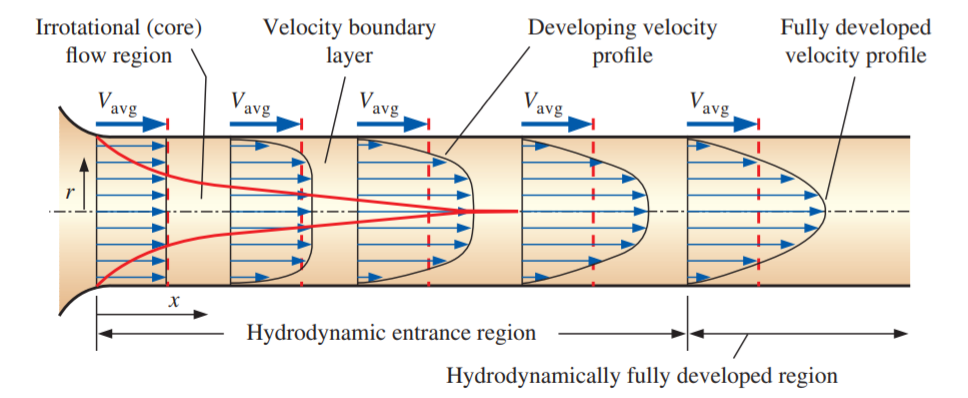}
\caption{Hydrodynamic entry length}
\end{figure}
\subsubsection{Multiphase CFD model}
Our model describes blood as a multiphase system wherein two dispersed phases, RBCs and leukocytes are suspended in plasma. Our model is an extension of the multiphase CFD approach proposed in Gidaspow\cite{gidaspow1994multiphase} and Anderson and Jackson\cite{anderson1967fluid}. This approach is particularly successful in describing blood flows with closely spaced RBCs, which have a volume fraction in the range 30-55 percent in vivo\cite{jung2006multiphase}. Compared to the single phase model, the multiphase model incorporates the volume fraction of each phase as well as mechanisms for exchange of momentum between the phases.

\subsubsection{Continuity equations}
The continuity equation for each phase (k = plasma, RBCs, leukocytes) is given by

\subsubsection{Solver and Boundary conditions} 
To obtain the numerical solution of above non-linear coupled diffrential equations, we use the commercial software package ANSYS FLUENT, which has a 3D Eulerian multiphase model built into it, without custom equations for interparticle phenomena. We use our own equations to write user defined functions for shear thinning viscosity, drag force and transient inlet-mixture velocity waveforms. Two external forces of virtual mass and lift force, being very small as compared to the drag force, were neglected. The numerical solution method uses an implicit, finite volume, structured mesh. A volume fraction correction equation in the continuity equations was used for numerical convergence. The boundary conditions are velocity inlet, pressure outlet, walls. Considering the symmetrical geometry about the axis, an axis boundary condition is also imposed. For initial conditions, the velocities were always set to zero. Subsequently, equal inlet velocities were given to all three phases(two phases for the two phase validation case).

\subsection{Validation with experimental results}
When we applied our multiphase CFD model including the constitutive relations in order to estimate the hemodynamics in disturbed flow regions, the two phase dilute flow simulation of hardened human red cells in steady-state condition agreed well with experimental results in the vortex formations, reattachment points and velocity profiles. Further, the velocities on the PQ line of the cross section through the vortex centre showed the best match to the measured velocities. The velocities at all points indicated in Figure 1 agreed within the same order of magnitude, as shown in Table 2.
\begin{figure}[h!]
\includegraphics{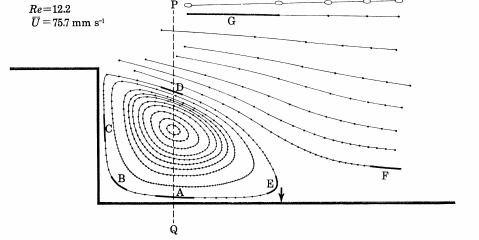}
\caption{Line PQ and other points of consideration}
\end{figure}
\begin{figure}[h!]
\includegraphics[scale=0.4]{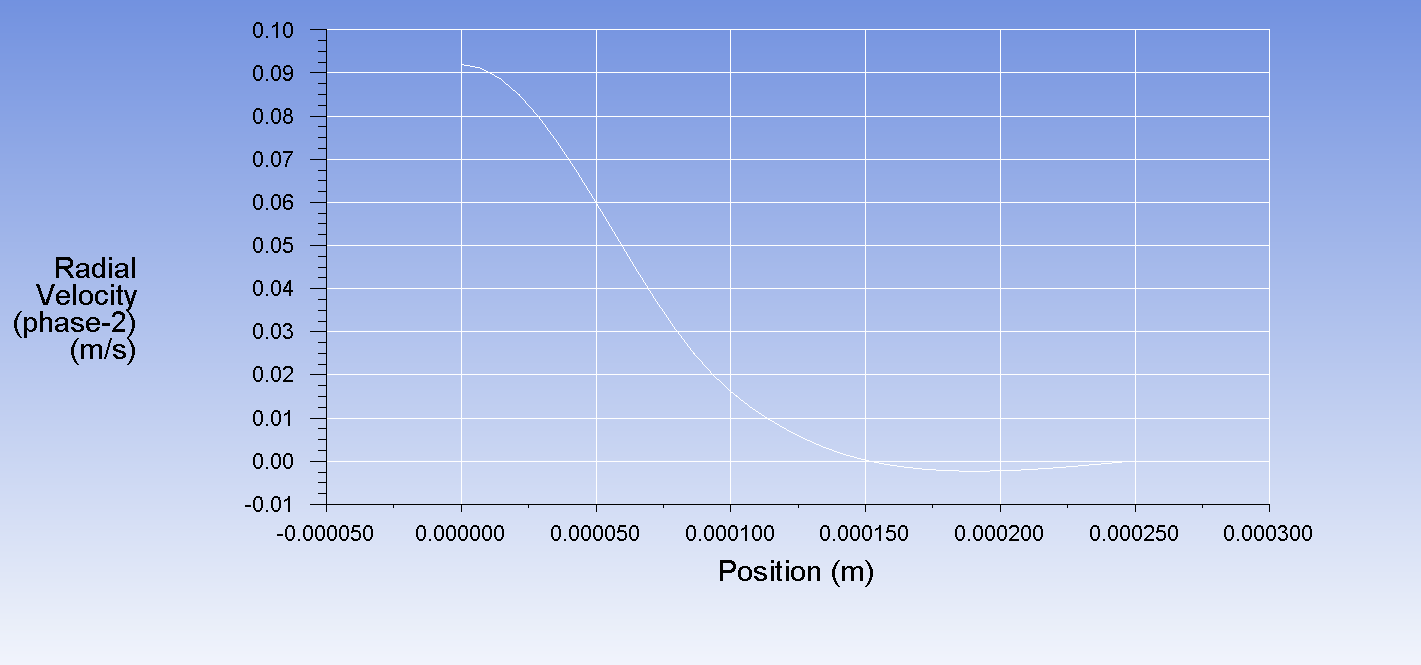}
\caption{Velocity of phase 2(RBC) along line PQ, drawn through vortex centre}
\end{figure}
\begin{figure}[h!]
  \includegraphics[scale=0.25]{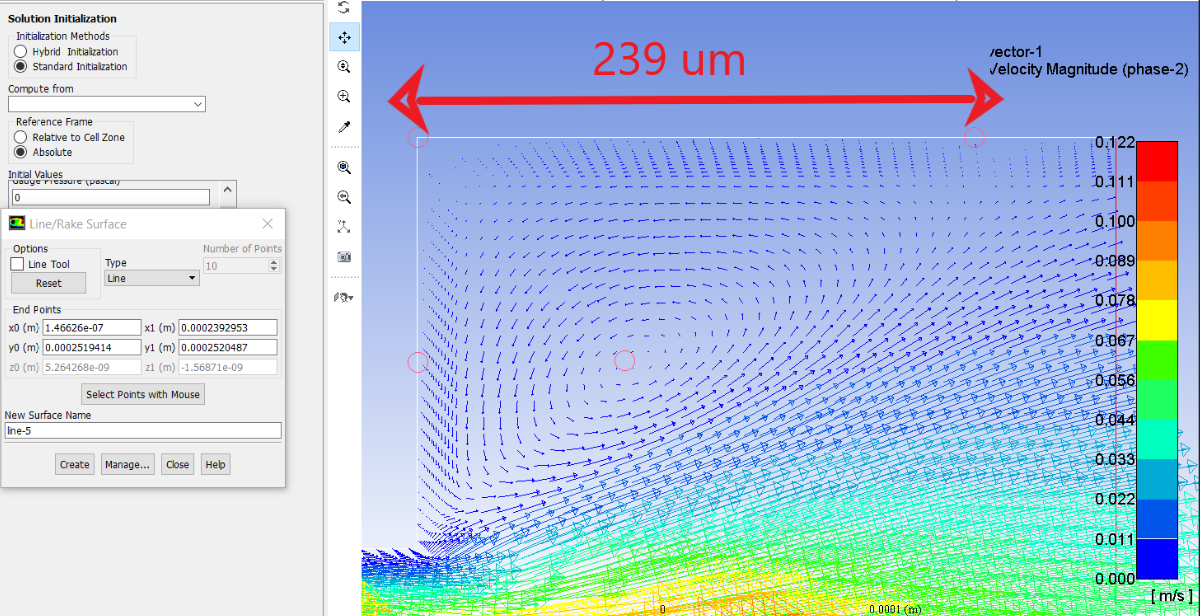}
  \includegraphics[scale=0.25]{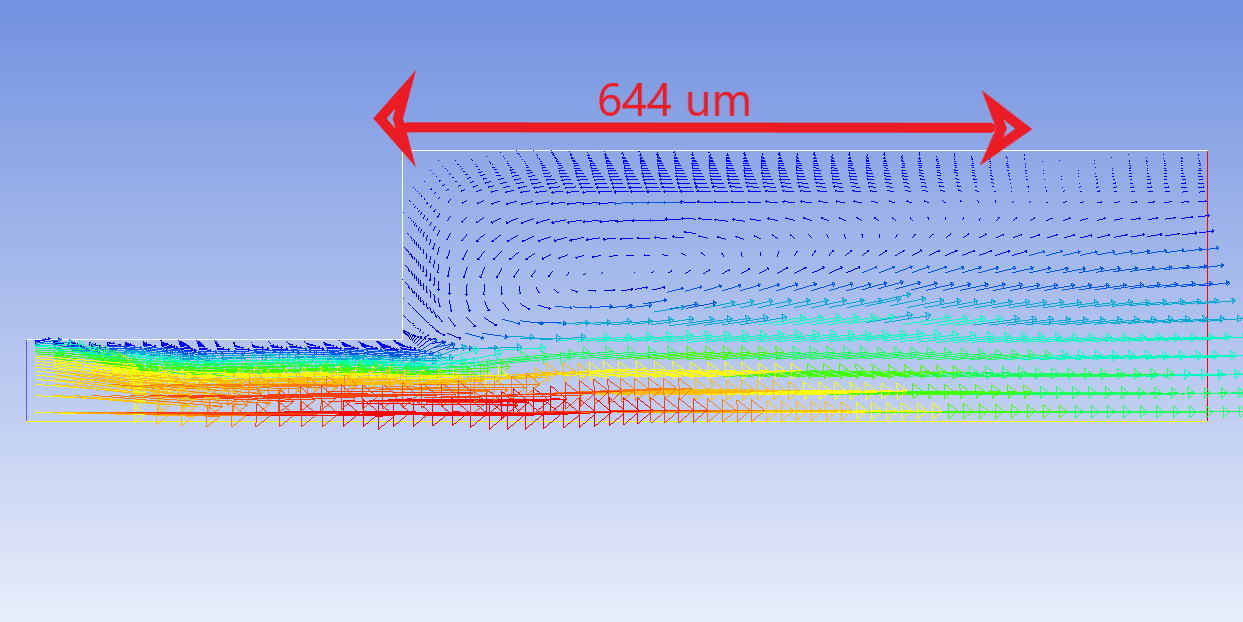}
  \caption{Reattachment length for Re = 12.2(left) and Re = 37.8(right)}
\end{figure}
\begin{table}
\caption{\label{arttype}Mesh parameters}
\footnotesize\rm
\begin{tabular*}{\textwidth}{@{}l*{15}{@{\extracolsep{0pt plus12pt}}l}}
Parameter&Recommended Range&Value from our mesh\\
\verb"Aspect ratio"&$<$20&1.2\\
\verb"Orthogonal Quality"&$\approx$1&0.96\\
\verb"Skewness"&$<$0.85&0.21\\
\end{tabular*}
\end{table}

\begin{table}
\caption{\label{arttype}Comparision of computational and experimental particle velocity magnitude(mm/s) in vortex regions, for Re = 12.2}
\footnotesize\rm
\begin{tabular*}{\textwidth}{@{}l*{15}{@{\extracolsep{0pt plus12pt}}l}}
Location&Experiment&Computation\\
\verb"O"&0.54&0.77\\
\verb"A"&0.64&2.03\\
\verb"B"&0.36&0.4\\
\verb"C"&1.46&2.09\\
\verb"D"&11.8&18.15\\
\verb"E"&0.3&0.56\\
\verb"F"&3.3&1.304\\
\verb"G"&48.6&64.01\\
\end{tabular*}
\end{table}

\section{Part 2: Multiphase CFD model applied to tumor vasculature}
\subsection{Geometry of artery with tumor vasculature}
Typically any blood vessel with a tumor growth looks like the one shown in Figure 4. For pancreatic tumor too, there is blood vessel growth near the tumor, which offers a possibility of particulate transport into the tumor vasculature\cite{wijeratne2017plos}.For our numerical simulation, we simplify the geometry to include only one main artery with pressure driven flow, and another branching out to the tumor region. The tumor vasculature is assumed to be stacked cylinders, for simplicity. This is shown in Figure 5. Blood flows from pressure P1 to P2 and possibly enters into the tumor region with interstitial pressure P3, through a hole. Note that P1, P2, P3 are related as P1$<$P2$<$P3.
\begin{figure}[h!]
  \includegraphics{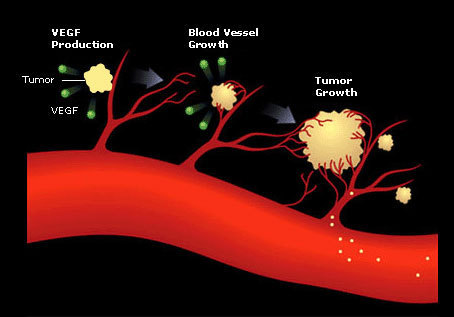}
  \caption{Blood vessel with tumor}
\end{figure}
\begin{figure}[h!]
  \includegraphics[scale=0.4]{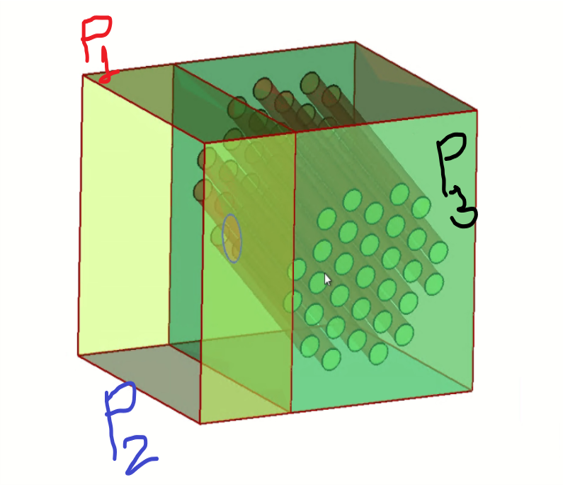}
  \includegraphics[scale=0.35]{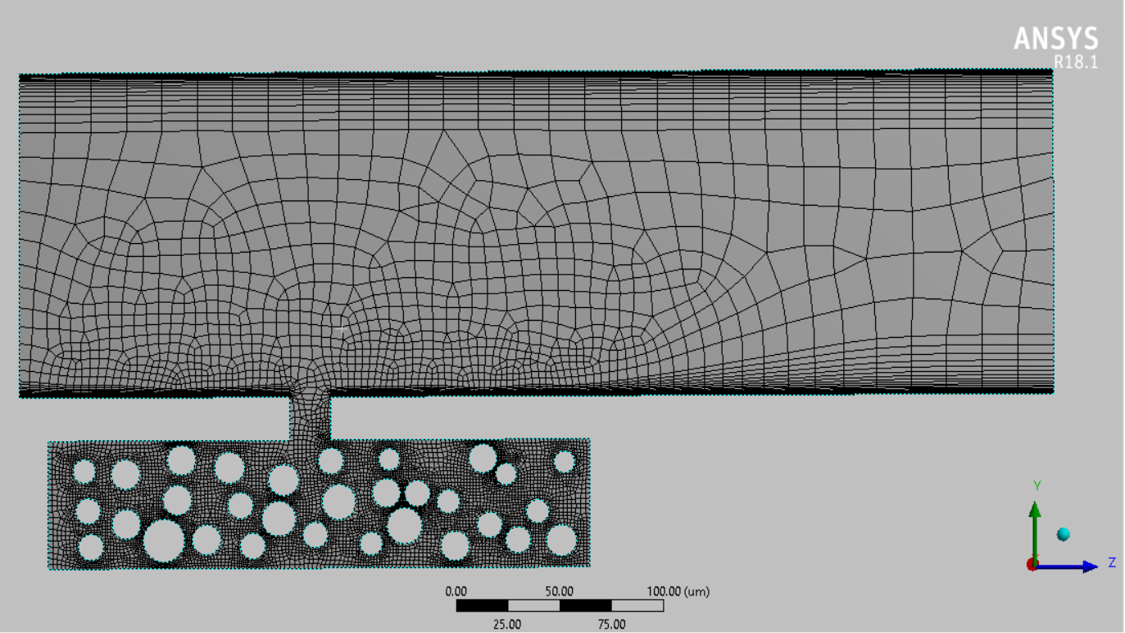}
  \caption{Simplified tumor vasculature(left) and mesh in section plane(right)}
\end{figure}
\subsection{Boundary and initial conditions}
All simulations are carried out in two dimensional section plane in order to minimize computational resources on a personal computer. Initially we assume that the artery is filled with only plasma and RBCs, and the tumor region is empty,i.e., filled with air(Figure 6). The inlet, outlet and interstitial pressures are 10 Pa, 5 Pa and 7.5 Pa respectively. The inlet velocity is pulsatile with a time period of 0.735 s(Figure 7). To match it with an actual heart cycle, experimental data\cite{sefidgar2015cmmm} is mapped into a Fourier transform in MATLAB curve fitting tool. Figure 7 also shows how the mapped inlet velocity function is periodic.
\begin{figure}[h!]
  \includegraphics[scale=0.33]{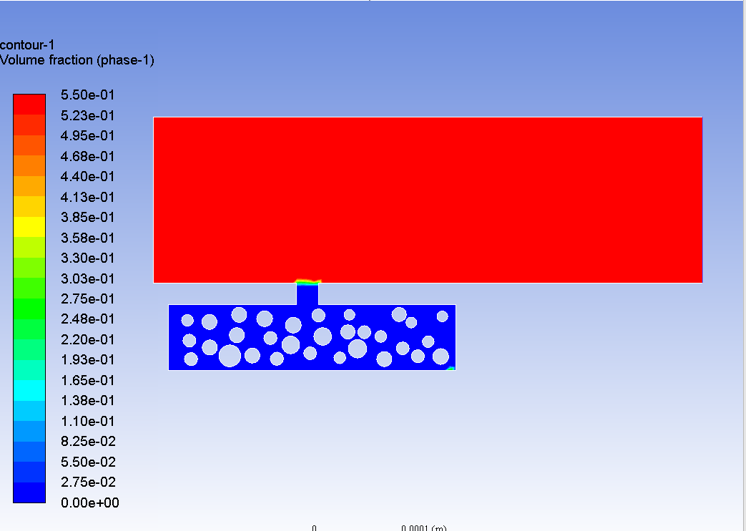}
  \includegraphics[scale=0.32]{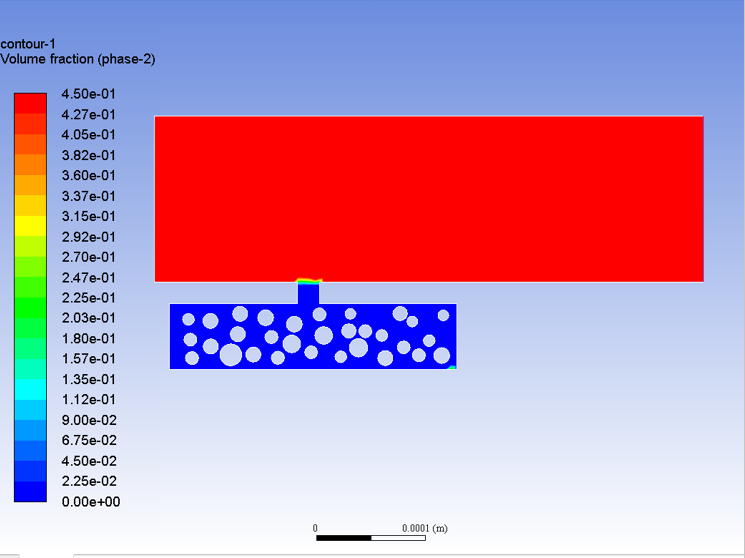}
  \includegraphics[scale=0.33]{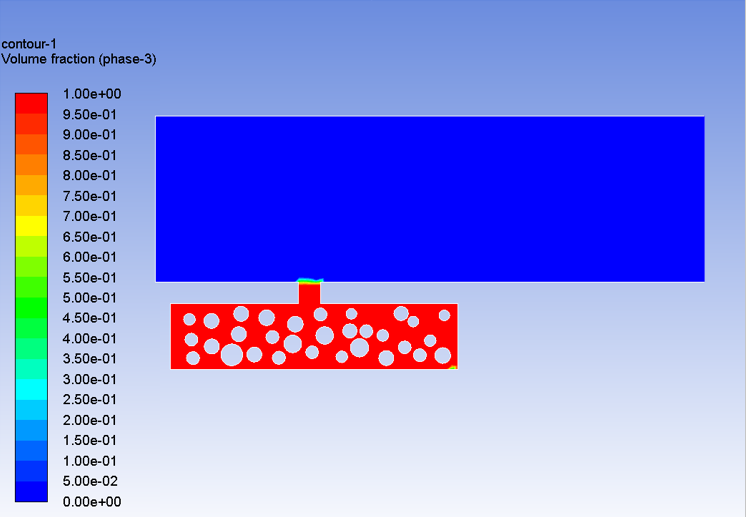}
  \caption{Initial conditions: (a) Volume fraction of plasma (b) Volume fraction of RBCs (c) Volume fraction of air }
\end{figure}
\begin{figure}[h!]
  \includegraphics[scale=0.6]{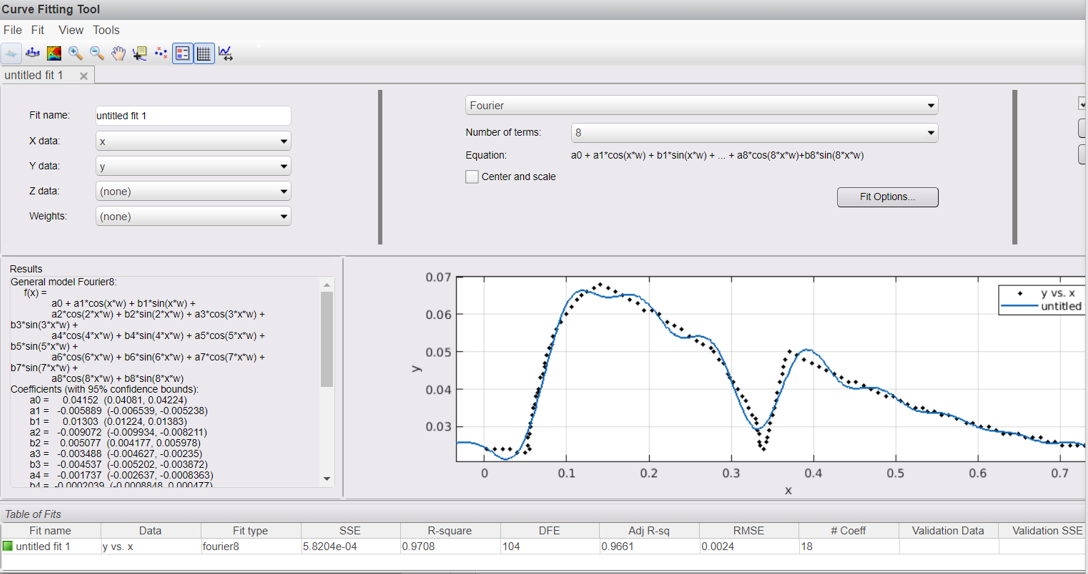}
  \includegraphics[scale=0.6]{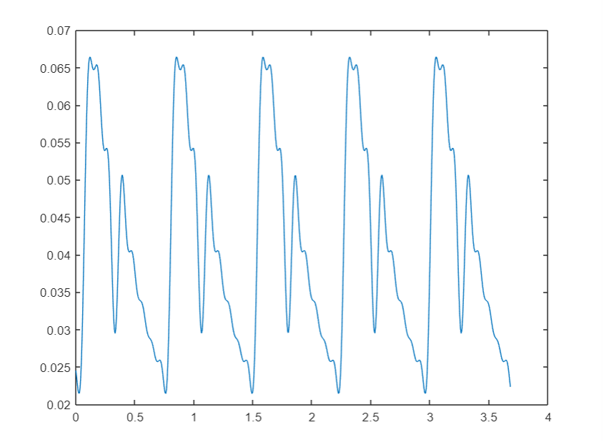}
  \caption{(a) Mapped inlet velocity function (b) Periodicity of mapped function}
\end{figure}
\begin{figure}[h!]
  \includegraphics[scale=0.6]{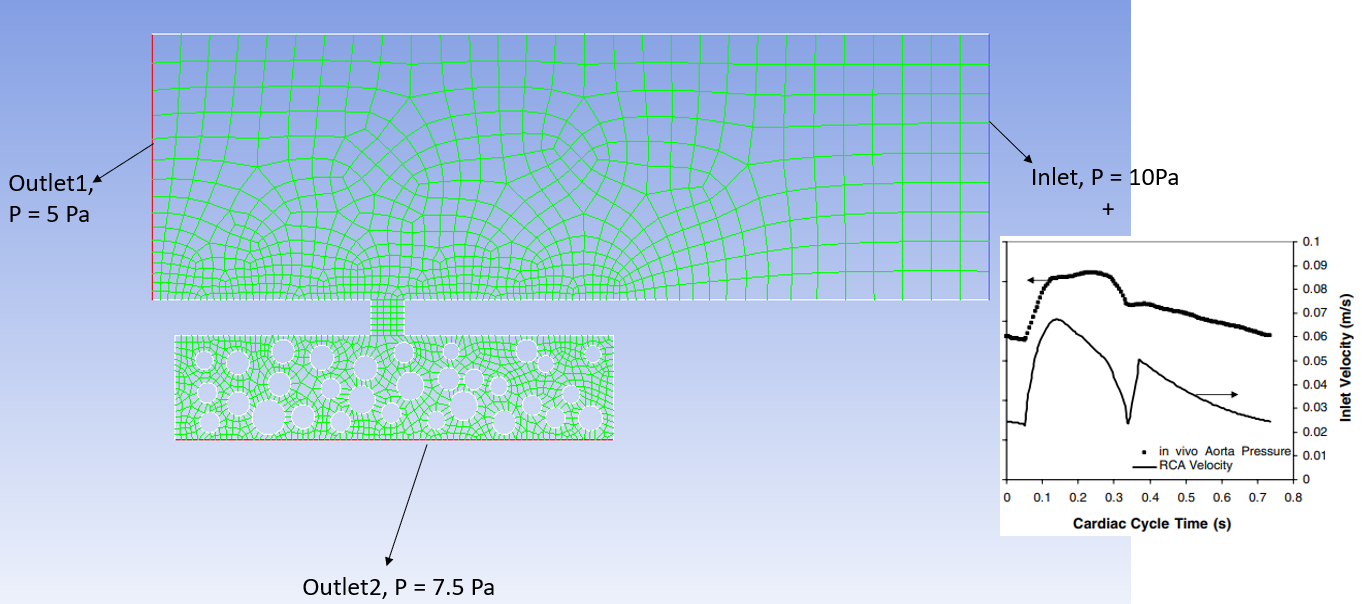}
  \caption{Boundary conditions}
\end{figure}

\subsection{Results}
The simulations are run for two heart cycles for a total time of two hours, on a 1.60 GHz personal computer. The velocity vectors of RBCs are shown in Figure 9. While relatively less number of RBCs can penetrate into the tumor region, it is clear from the volume fraction contours of RBCs that the final particulate distribution is a function of the geometry of the tumor(Figure 10).
\begin{figure}[h!]
  \includegraphics[scale=0.5]{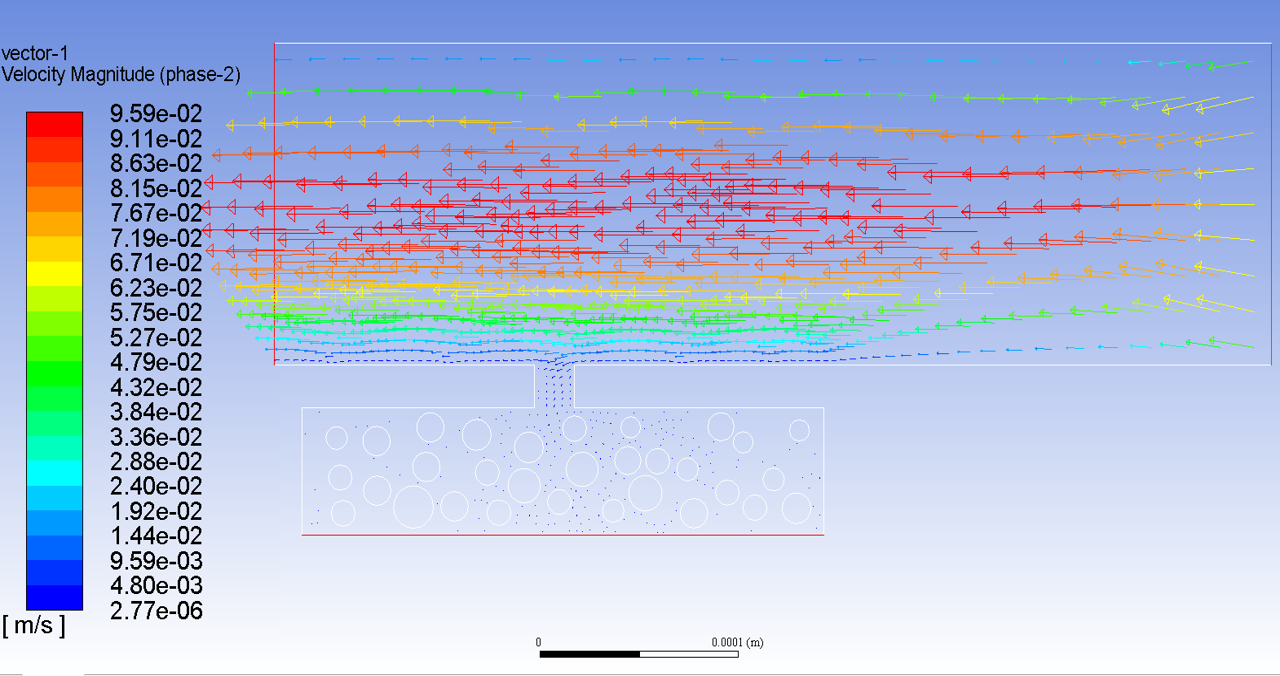}
  \caption{Velocity vectors of RBC after two heart cycles}
\end{figure}
\begin{figure}[h!]
  \includegraphics[scale=0.5]{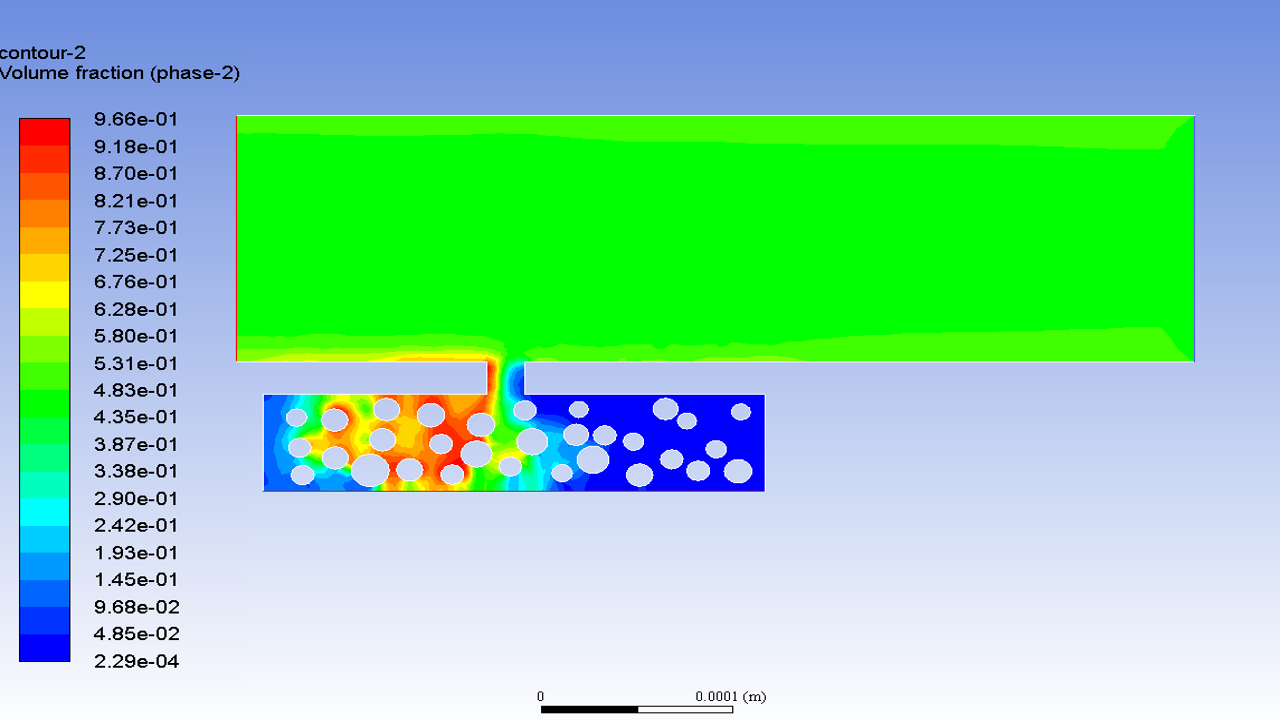}
  \caption{RBC volume fraction after two heart cycles}
\end{figure}
\bibliography{CBM_references}

\end{document}